\documentclass[aps,prd,preprint,tightenlines,superscriptaddress,showpacs,byrevtex]{revtex4}
\usepackage{graphicx}
\usepackage{dcolumn}
\usepackage{bm}
\usepackage{pstricks}
\usepackage{pst-node}

\begin{document}

\vspace*{-3\baselineskip}
\resizebox{!}{3cm}{\includegraphics{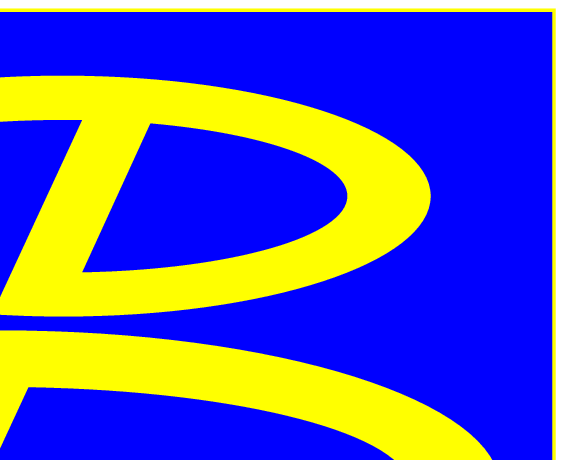}}
\preprint{\vbox{
		  \hbox{KEK   Preprint 2004-105}
		  \hbox{Belle Preprint 2005-9}
}}

\def\bz{{B^0}}
\def\bzb{{\overline{B}{}^0}}
\def\bp{{B^+}}
\def\bm{{B^-}}
\def\bpm{{B^{\pm}}}
\def\dE{{\Delta E}}
\def\mb{{M_{\rm bc}}}
\def\Dt{\Delta t}
\def\Dz{\Delta z}
\def\fol{f_{\rm ol}}
\def\fsig{f_{\rm sig}}
\newcommand{\sinbb}{{\sin2\phi_1}}
 
\newcommand{\ra}{\rightarrow}
\newcommand{\myindent}{\hspace*{2cm}}  
\newcommand{\fCP}{f_{CP}}
\def\fcp{\fCP}
\newcommand{\ftag}{f_{\rm tag}}
\newcommand{\zCP}{z_{CP}}
\newcommand{\tCP}{t_{CP}}
\newcommand{\ttag}{t_{\rm tag}}
\newcommand{\cala}{{\cal A}}
\newcommand{\calb}{{\cal B}}
\newcommand{\cals}{{\cal S}}
\newcommand{\dm}{\Delta m_d}
\newcommand{\dmd}{\dm}
\def\taubz{{\tau_\bz}}
\def\taubp{{\tau_\bp}}
\def\ks{{K_S^0}}
\def\kl{K_L^0}
\def\kpm{{K^{\pm}}}
\newcommand{\btoccs}{b \to c\overline{c}s}
\newcommand{\btosqq}{b \to s\overline{q}q}
\newcommand{\btosss}{b \to s\overline{s}s}
\newcommand*{\dwl}{\ensuremath{{\Delta w_l}}}
\newcommand*{\fq}{\ensuremath{q}}
\def\kz{{K^0}}
\def\kp{{K^+}}
\def\km{{K^-}}
\def\fzero{{f_0(980)}}
\def\pip{{\pi^+}}
\def\pim{{\pi^-}}
\def\piz{{\pi^0}}
\def\bbar{{\overline{B}}}
\def\ufs{{\Upsilon(4S)}}
\def\nev{{N_{\rm ev}}}
\def\nsig{{N_{\rm sig}}}
\def\Nev{\nev}
\def\nsigmc{{N_{\rm sig}^{\rm MC}}}
\def\nbkg{{N_{\rm bkg}}}

\def\jpsi{{J/\psi}}
\def\dzb{{\overline{D}{}^0}}
\newcommand{\dslnu}{D^{*-}\ell^+\nu}
\newcommand{\bzdslnu}{\bz \to \dslnu}

\def\lsig{{\cal L}_{\rm sig}}
\def\lbkg{{\cal L}_{\rm bkg}}
\def\rsigbkg{{\cal R}_{\rm s/b}}
\def\calf{{\cal F}}

\def\mgg{M_{\gamma\gamma}}
\def\ppizcms{p_\piz^{\rm cms}}

\def\pbstar{p_B^{\rm cms}}

\newcommand*{\eeff}{\ensuremath{\epsilon_\textrm{eff}}}
\def\egcms{{E_\gamma^{\rm cms}}}

\def\lnsig{{\cal L}_{N_{\rm sig}}}
\def\lzero{{\cal L}_0}

\def\acpraw{{A_{CP}^{\rm raw}}}

\def\rsig{R_{\rm sig}}

\def\kspm{{K_S^{+-}}}
\def\kszz{{K_S^{00}}}

\def\efftot{{0.30\pm 0.01}}

\def\sinbbWA{+0.726}
\def\sinbbERR{0.037}
\def\sinbbWAResult{\sinbbWA\pm\sinbbERR}


\def\NBevksksks{167} 
\def\PBksksks{0.53} \def\NBsigksksks{88\pm 13}

\def\NBevkspmkspmkspm{128} 
\def\PBkspmkspmkspm{0.56} \def\NBsigkspmkspmkspm{72\pm 10}

\def\NBevkspmkspmkszz{39} 
\def\PBkspmkspmkszz{0.40} \def\NBsigkspmkspmkszz{16\pm 8}

\def\NAevksksks{117} 
\def\PAksksks{0.54} \def\NAsigksksks{63\pm 9}

\def\NAevkspmkspmkspm{96} 
\def\PAkspmkspmkspm{0.56} \def\NAsigkspmkspmkspm{54\pm 8}

\def\NAevkspmkspmkszz{21} 
\def\PAkspmkspmkszz{0.40} \def\NAsigkspmkspmkszz{8\pm 4}

\def\SksksksVal{+1.26} \def\SksksksStat{0.68} \def\SksksksSyst{0.20}
\def\AksksksVal{+0.54} \def\AksksksStat{0.34} \def\AksksksSyst{0.09}

\def\SksksksResult{\SksksksVal\pm\SksksksStat\pm\SksksksSyst}
\def\SksksksResultSS
  {\SksksksVal\pm\SksksksStat\mbox{(stat)}\pm\SksksksSyst\mbox{(syst)}}

\def\AksksksResult{\AksksksVal\pm\AksksksStat\pm\AksksksSyst}
\def\AksksksResultSS
  {\AksksksVal\pm\AksksksStat\mbox{(stat)}\pm\AksksksSyst\mbox{(syst)}}

\def\SkskskpmVal{+0.33} \def\SkskskpmStat{^{+0.39}_{-0.45}}
\def\AkskskpmVal{-0.32} \def\AkskskpmStat{0.28}

\def\SkskskpmResultS
  {\SkskskpmVal\SkskskpmStat\mbox{(stat)}}

\def\AkskskpmResultS
  {\AkskskpmVal\pm\AkskskpmStat\mbox{(stat)}}

\def\SbsqqNewVal{+0.39} \def\SbsqqNewErr{0.11}
\def\SbsqqNewResult{\SbsqqNewVal\pm\SbsqqNewErr}

\def\TauksksksVal{0.91} \def\TauksksksStat{^{+0.29}_{-0.23}} 
\def\TauksksksResultS{\TauksksksVal\TauksksksStat\mbox{(stat)}}

\def\TaukskskpmVal{1.29} \def\TaukskskpmStat{^{+0.27}_{-0.22}} 
\def\TaukskskpmResultS{\TaukskskpmVal\TaukskskpmStat\mbox{(stat)}}

\def\SbsqqVal{+0.43} \def\SbsqqErr{^{+0.12}_{-0.11}}
\def\SbsqqResult{\SbsqqVal\SbsqqErr}
\def\BRksksks{(4.2^{+1.6}_{-1.3}\pm{0.8})~\times~10^{-6}}

\title{\quad\\[0.5cm] \boldmath Measurement of Time-Dependent 
{\boldmath $CP$}-Violating\\ Asymmetries
in $\bz\to\ks\ks\ks$ Decay}

\date{\today}

\affiliation{Budker Institute of Nuclear Physics, Novosibirsk}
\affiliation{Chiba University, Chiba}
\affiliation{Chonnam National University, Kwangju}
\affiliation{University of Cincinnati, Cincinnati, Ohio 45221}
\affiliation{University of Frankfurt, Frankfurt}
\affiliation{University of Hawaii, Honolulu, Hawaii 96822}
\affiliation{High Energy Accelerator Research Organization (KEK), Tsukuba}
\affiliation{Hiroshima Institute of Technology, Hiroshima}
\affiliation{Institute of High Energy Physics, Chinese Academy of Sciences, Beijing}
\affiliation{Institute of High Energy Physics, Vienna}
\affiliation{Institute for Theoretical and Experimental Physics, Moscow}
\affiliation{J. Stefan Institute, Ljubljana}
\affiliation{Kanagawa University, Yokohama}
\affiliation{Korea University, Seoul}
\affiliation{Kyungpook National University, Taegu}
\affiliation{Swiss Federal Institute of Technology of Lausanne, EPFL, Lausanne}
\affiliation{University of Ljubljana, Ljubljana}
\affiliation{University of Maribor, Maribor}
\affiliation{University of Melbourne, Victoria}
\affiliation{Nagoya University, Nagoya}
\affiliation{Nara Women's University, Nara}
\affiliation{National Central University, Chung-li}
\affiliation{National United University, Miao Li}
\affiliation{Department of Physics, National Taiwan University, Taipei}
\affiliation{H. Niewodniczanski Institute of Nuclear Physics, Krakow}
\affiliation{Nihon Dental College, Niigata}
\affiliation{Niigata University, Niigata}
\affiliation{Osaka City University, Osaka}
\affiliation{Osaka University, Osaka}
\affiliation{Panjab University, Chandigarh}
\affiliation{Peking University, Beijing}
\affiliation{Princeton University, Princeton, New Jersey 08544}
\affiliation{Saga University, Saga}
\affiliation{University of Science and Technology of China, Hefei}
\affiliation{Seoul National University, Seoul}
\affiliation{Sungkyunkwan University, Suwon}
\affiliation{University of Sydney, Sydney NSW}
\affiliation{Tata Institute of Fundamental Research, Bombay}
\affiliation{Toho University, Funabashi}
\affiliation{Tohoku Gakuin University, Tagajo}
\affiliation{Tohoku University, Sendai}
\affiliation{Department of Physics, University of Tokyo, Tokyo}
\affiliation{Tokyo Institute of Technology, Tokyo}
\affiliation{Tokyo Metropolitan University, Tokyo}
\affiliation{Tokyo University of Agriculture and Technology, Tokyo}
\affiliation{University of Tsukuba, Tsukuba}
\affiliation{Virginia Polytechnic Institute and State University, Blacksburg, Virginia 24061}
\affiliation{Yonsei University, Seoul}
   \author{K.~Sumisawa}\affiliation{Osaka University, Osaka}\affiliation{High Energy Accelerator Research Organization (KEK), Tsukuba} 
   \author{Y.~Ushiroda}\affiliation{High Energy Accelerator Research Organization (KEK), Tsukuba} 
   \author{M.~Hazumi}\affiliation{High Energy Accelerator Research Organization (KEK), Tsukuba} 
   \author{K.~Abe}\affiliation{High Energy Accelerator Research Organization (KEK), Tsukuba} 
   \author{K.~Abe}\affiliation{Tohoku Gakuin University, Tagajo} 
   \author{I.~Adachi}\affiliation{High Energy Accelerator Research Organization (KEK), Tsukuba} 
   \author{H.~Aihara}\affiliation{Department of Physics, University of Tokyo, Tokyo} 
   \author{Y.~Asano}\affiliation{University of Tsukuba, Tsukuba} 
   \author{V.~Aulchenko}\affiliation{Budker Institute of Nuclear Physics, Novosibirsk} 
   \author{T.~Aushev}\affiliation{Institute for Theoretical and Experimental Physics, Moscow} 
% ...
\collaboration{The Belle Collaboration}

\begin{abstract}
  We present a measurement of $CP$-violation parameters
  in the
  $\bz\to\ks\ks\ks$
  decay
  based on a sample of $275\times 10^6$ $B\bbar$ pairs
  collected at the $\ufs$ resonance with
  the Belle detector at the KEKB energy-asymmetric $e^+e^-$ collider.
  One neutral $B$ meson is fully reconstructed in
  the decay $\bz\to\ks\ks\ks$,
  and the flavor of the accompanying $B$ meson is identified from
  its decay products.
  $CP$-violation parameters
  are obtained from the asymmetry in the distributions of
  the proper-time interval between the two $B$ decays:
 $\cals = \SksksksResultSS$ and $\cala = \AksksksResultSS$.
\end{abstract}


\maketitle


In the Standard Model (SM), $CP$ violation arises from
the Kobayashi-Maskawa phase~\cite{Kobayashi:1973fv}
in the weak-interaction quark-mixing matrix.
In particular, 
the SM predicts to a good approximation that 
$\cals = -\xi_f\sin 2\phi_1$ and $\cala =0$
for both $\btoccs$ and $\btosqq$ transitions,
where $\cals$ ($\cala$) is a parameter for
mixing-induced (direct) $CP$ violation~\cite{bib:sanda},
$\xi_f = +1 (-1)$ corresponds to
$CP$-even (-odd) final states,
and $\phi_1$ is one of angles of the Unitarity Triangle.
Measurements of time-dependent $CP$ asymmetries in
$\bz \to J/\psi \ks$~\cite{bib:CC} and related decay 
modes, which are governed by the $b \to c\overline{c}s$ transition,
by the Belle~\cite{bib:CP1_Belle,bib:BELLE-CONF-0436}
and BaBar~\cite{bib:CP1_BaBar} collaborations
already determine $\sinbb$ rather precisely;
the present world average value is 
$\sinbb = \sinbbWAResult$~\cite{bib:HFAG}.

$CP$-violation parameters in the flavor-changing $b \to s$ transition
are sensitive to phenomena at a very high-energy
scale~\cite{bib:lucy,Akeroyd:2004mj}.
Belle measurements~\cite{Abe:2004xp}
in the decay modes $\bz\to$ $\phi\ks$, $\phi\kl$, $\kp\km\ks$,
$\fzero\ks$, $\eta'\ks$, $\omega\ks$, and $\ks\piz$, which are dominated
by the $\btosqq$ transition, yield $\sinbb = \SbsqqResult$
when all the modes are combined.
Measurements by the BaBar collaboration
also yield a similar deviation~\cite{bib:HFAG, bib:BaBar_sss}.
To elucidate the difference in $CP$ asymmetries
between $\btosqq$ and $\btoccs$ transitions,
it is essential to examine additional modes that
may be sensitive to the same $b \to s$ amplitude.

The $\bz$ decay to $\ks\ks\ks$, which is a $\xi_f = +1$ state,
is one of the most promising
modes for this purpose~\cite{Gershon:2004tk}.
Since there is no $u$ quark in the final state,
the decay is dominated by the $\btosss$ transition.
Its branching fraction $\calb(\bz\to\ks\ks\ks) = \BRksksks$ was reported
by Belle~\cite{Garmash:2003er}.
In this Letter, 
we describe a measurement of
$CP$ asymmetries in the $\bz\to\ks\ks\ks$ decay.

In the decay chain $\Upsilon(4S)\to \bz\bzb \to f_{CP}f_{\rm tag}$,
where one of the $B$ mesons decays at time $t_{CP}$ to  
a $CP$ eigenstate $f_{CP}$ 
and the other decays at time $t_{\rm tag}$ to a final state  
$f_{\rm tag}$ that distinguishes between $B^0$ and $\bzb$, 
the decay rate has a time dependence
given by
\begin{eqnarray}
\label{eq:psig}
{\cal P}(\Delta{t}) = 
\frac{e^{-|\Delta{t}|/{\taubz}}}{4{\taubz}}
\biggl\{1 + \fq\cdot 
\Bigl[ \cals\sin(\dmd\Delta{t}) \nonumber \\
   + \cala\cos(\dmd\Delta{t})
\Bigr]
\biggr\}.
\end{eqnarray}
Here $\taubz$ is the $B^0$ lifetime, $\dmd$ is the mass difference 
between the two $B^0$ mass eigenstates,
$\Delta{t}$ is the time difference $t_{CP}$ $-$ $t_{\rm tag}$, and
the $b$-flavor charge is $\fq$ = +1 ($-1$) when the tagging $B$ meson
is a $B^0$ ($\bzb$).

At the KEKB energy-asymmetric 
$e^+e^-$ (3.5 on 8.0~GeV) collider~\cite{bib:KEKB},
the $\Upsilon(4S)$ resonance is produced
with a Lorentz boost of $\beta\gamma=0.425$
along the $z$ direction which is antiparallel to the positron beamline.
Since the $B^0$ and $\bzb$ mesons are approximately at 
rest in the $\Upsilon(4S)$ center-of-mass system (cms),
$\Delta t$ can be determined from the displacement in $z$ 
between the $f_{CP}$ and $f_{\rm tag}$ decay vertices:
$\Delta t \simeq (z_{CP} - z_{\rm tag})/(\beta\gamma c)
 \equiv \Delta z/(\beta\gamma c)$.

The Belle detector~\cite{Belle} is a large-solid-angle magnetic
spectrometer that consists of a silicon vertex detector (SVD), a
50-layer central drift chamber, an array of aerogel threshold
Cherenkov counters, a barrel-like arrangement of time-of-flight
scintillation counters, an electromagnetic calorimeter (ECL) and an
iron flux-return instrumented to detect $K_L^0$ mesons and to identify
muons.  A 2.0\,cm radius beampipe and a 3-layer SVD (SVD-I) were used
for a 140 fb$^{-1}$ data sample,
while a 1.5 cm radius beampipe, a 4-layer
silicon detector (SVD-II)~\cite{Ushiroda} 
and a small-cell inner drift chamber were used for 
an additional 113 fb$^{-1}$ data sample. 
In total $275\times 10^6$ $B\bbar$ pairs were accumulated.

We reconstruct the $\bz\to\ks\ks\ks$ decay
in the $\kspm\kspm\kspm$ or $\kspm\kspm\kszz$ final state,
where the $\pip\pim$ ($\piz\piz$) 
state from a $\ks$ decay is denoted as $\kspm$ ($\kszz$).
%
%
Pairs of oppositely charged tracks with the $\pip\pim$ invariant mass
within 0.012~GeV/$c^2$ ($\simeq 3 \sigma$)
of the nominal $\ks$ mass are
used to reconstruct $\kspm$ candidates.
The $\pip\pim$ vertex is required to be displaced from
the interaction point (IP) by a minimum transverse distance of 0.22~cm
for candidates with p $>1.5$~GeV/$c$ and 0.08~cm for
those with p $<1.5$~GeV/$c$,
where p is the momentum of $\ks$.
The angle in the transverse plane between the $\ks$ momentum vector and the
direction defined by the $\ks$ vertex and the IP should be less than
0.03~rad (0.1~rad) for the high (low) momentum candidates.
The mismatch in the $z$ direction at the $\ks$ vertex point
for the two charged pion tracks should be less than 2.4~cm (1.8~cm)
for the high (low) momentum candidates.
After two good $\kspm$ candidates have been found which satisfy the criteria
given above, looser requirements are applied for the third $\kspm$ candidate.
The requirement on the transverse direction matching is relaxed to
0.2~rad (0.4~rad for low momentum candidates),
and the mismatch of the two charged pions in the $z$ direction
is required to be less than 5~cm (1~cm if both
pions have hits in the SVD).
To select $\kszz$ candidates,
we reconstruct $\piz$ candidates from pairs of photons
with $E_\gamma > 0.05$~GeV, where
$E_\gamma$ is the photon energy measured with the ECL.
The reconstructed $\piz$ candidate is required to have an invariant mass
between 0.08 and 0.15~GeV/$c^2$ and momentum above 0.1~GeV/$c$.
$\kszz $ candidates are required to have
an invariant mass between 0.47 and 0.52~GeV/$c^2$,
and a fit is performed with constraints on the
$\ks$ vertex and $\piz$ masses to improve the $\piz\piz$ invariant mass
resolution.
The $\kszz$ candidate is combined with two good $\kspm$ candidates
to reconstruct a $\bz$ meson.

To identify $\bz\to\ks\ks\ks$ decays, we use the
energy difference $\dE\equiv E_B^{\rm cms}-E_{\rm beam}^{\rm cms}$ and
the beam-energy constrained mass $\mb\equiv\sqrt{(E_{\rm beam}^{\rm cms})^2-
(p_B^{\rm cms})^2}$, where $E_{\rm beam}^{\rm cms}$ is
the beam energy in the cms, and
$E_B^{\rm cms}$ and $p_B^{\rm cms}$ are the cms energy and momentum of the 
reconstructed $B$ candidate, respectively.
The $\bz$ meson signal region is defined as 
$|\dE|<0.10$ GeV for $\bz \to \kspm\kspm\kspm$,
$-0.15 < \dE < 0.10$ GeV for $\bz \to \kspm\kspm\kszz$,
and $5.27 < \mb < 5.29~{\rm GeV}/c^2$ for both decays.
To suppress the $e^+e^- \rightarrow q\overline{q}$
continuum background ($q = u,~d,~s,~c$),
we form a signal over background likelihood ratio $\rsigbkg$ by
combining likelihoods for two quantities; a Fisher discriminant of
modified Fox-Wolfram moments~\cite{SFW}, and the cosine of the cms $\bz$
flight direction.
The requirement for $\rsigbkg$ depends both on the decay mode
and on the flavor-tagging quality;
after applying all other cuts,
this rejects 94\% of the $q\bar{q}$ background
while retaining 75\% of the signal.


If both $\bz\to\kspm\kspm\kspm$ and $\kspm\kspm\kszz$ candidates are found
in the same event, we choose the $\bz\to\kspm\kspm\kspm$ candidate.
When multiple $\bz\to\kspm\kspm\kspm$ candidates are found,
we prioritize those with three good $\kspm$ candidates.
If more than one candidate still remain,
we select the one with the smallest value for
$\sum(\Delta M_{\kspm})^2$, where $\Delta M_{\kspm}$ is the
difference between the reconstructed and nominal mass of $\kspm$.
For multiple $\bz\to\kspm\kspm\kszz$ candidates,
we select the $\kspm\kspm$ pair that has the smallest $\sum(\Delta
M_{\kspm})^2$ value and the $\kszz$ candidate with the minimum $\chi^2$
of the constrained fit.

We reject $\ks\ks\ks$ candidates if they are consistent with
$\bz\to\chi_{c0}\ks\to(\ks\ks)\ks$ or $\bz\to D^0\ks\to(\ks\ks)\ks$
decays, i.e. if one of the $\ks$ pairs has an invariant mass within
$\pm 2 \sigma$ of the $\chi_{c0}$ mass or $D^0$ mass, where $\sigma$ is
the $\ks\ks$ mass resolution.

\begin{figure}
\resizebox{0.49\columnwidth}{!}{\includegraphics{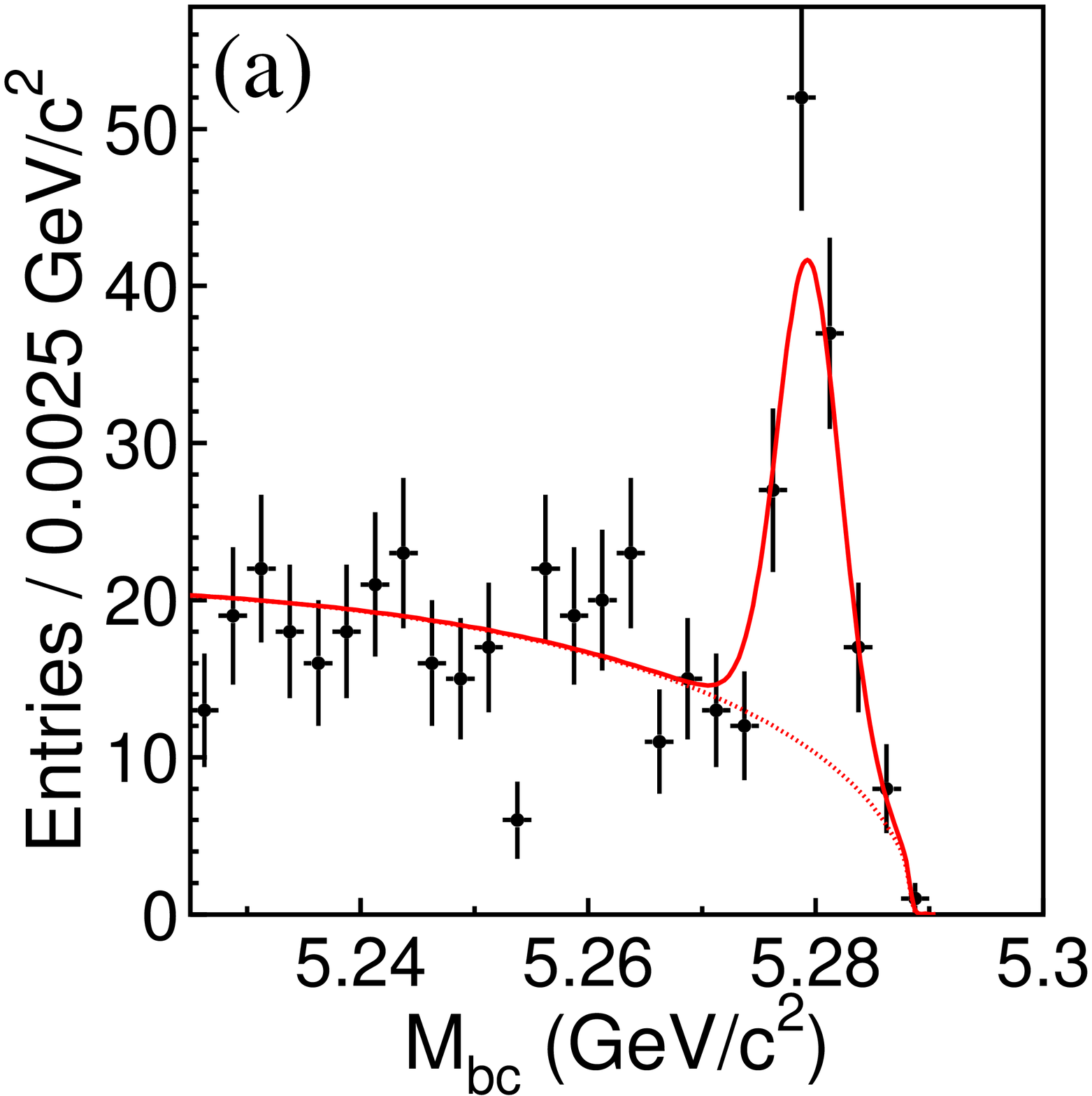}}
\resizebox{0.49\columnwidth}{!}{\includegraphics{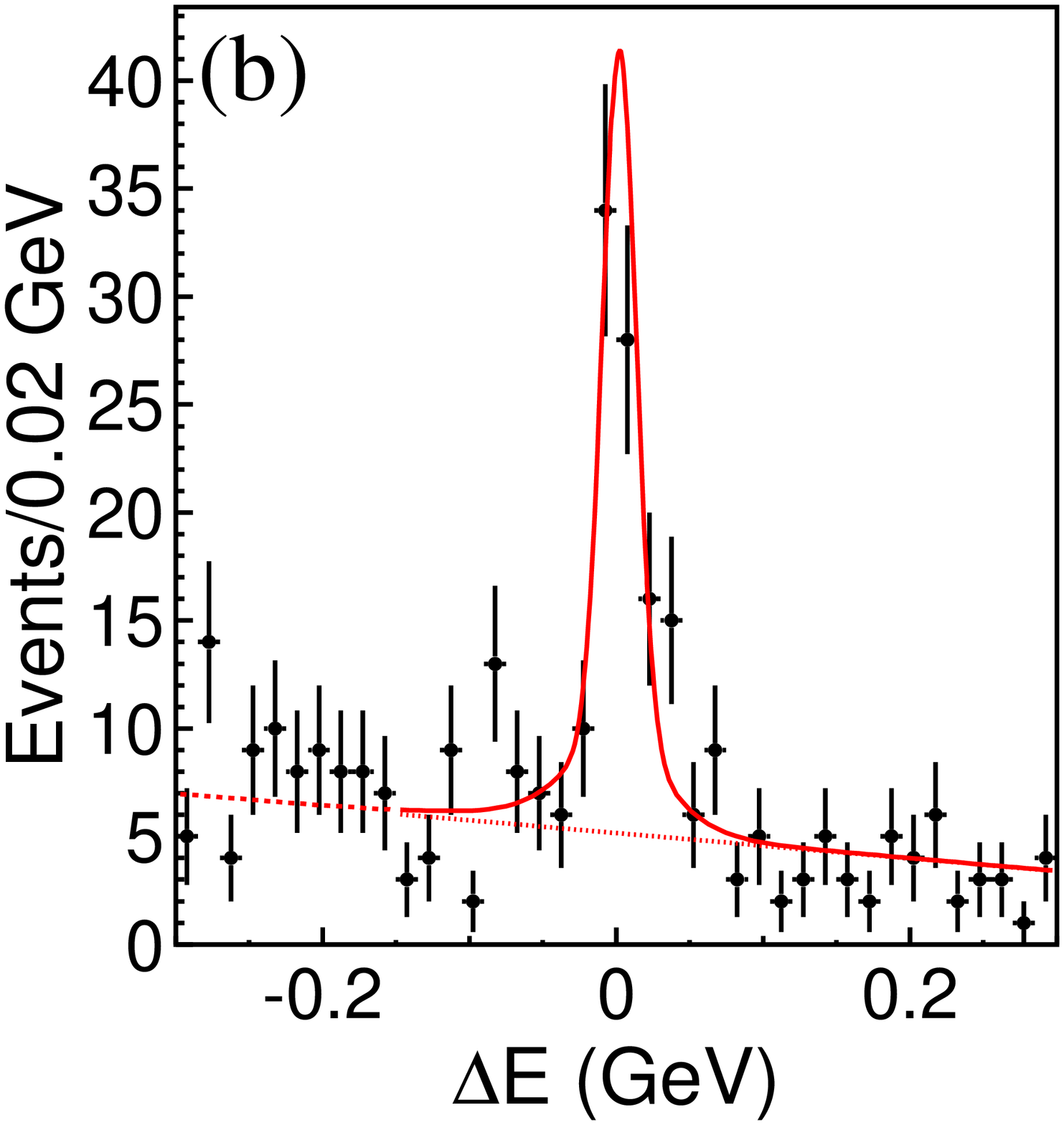}}
\caption{(a) $\mb$ distribution within the $\dE$ signal region,
(b) $\dE$ distribution within the $\mb$ signal region.
Solid curves show the fit to signal plus background distributions,
and dashed curves show the background contributions.}
\label{fig:mb}
\end{figure}

Figure~\ref{fig:mb} shows the $\mb$ and $\dE$ distributions for the
reconstructed $\bz\to\ks\ks\ks$ candidates
after flavor tagging.
The signal yield is determined
from an unbinned two-dimensional maximum-likelihood fit
to the $\dE$-$\mb$ distribution.
The $\kspm\kspm\kspm$ signal distribution 
is modeled with a Gaussian function (a sum of two Gaussian functions)
for $\mb$ ($\dE$).
For $\bz\to\kspm\kspm\kszz$ decay,
the signal is modeled with a two-dimensional
smoothed histogram obtained from Monte Carlo (MC) events.
For the continuum background,
we use the ARGUS parameterization~\cite{bib:ARGUS} 
for $\mb$ and a linear function for $\dE$.
The fits after flavor tagging yield 
$\NBsigkspmkspmkspm$ $\bz\to\kspm\kspm\kspm$ events and
$\NBsigkspmkspmkszz$ $\bz\to\kspm\kspm\kszz$ events
for a total of $\NBsigksksks$ $\bz\to\ks\ks\ks$ events 
in the signal region,
where the errors are statistical only.
The obtained purity is $\PBkspmkspmkspm$ for the $\kspm\kspm\kspm$
and $\PBkspmkspmkszz$ for the $\kspm\kspm\kszz$ channels.
We use events outside the signal region 
as well as a large MC sample to study the background components.
The dominant background is from continuum.
The contamination of $\bz\to\chi_{c0}\ks$ events in the 
$\bz\to\ks\ks\ks$ sample is small (less than $2.6$\% at 90\% C.L.).
The contributions from other $B\overline{B}$ events are negligibly small.
The influence of these backgrounds
is treated as a source of systematic uncertainty
in the $CP$ asymmetry measurement.
Backgrounds from the decay $\bz\to D^0\ks$
are found to be negligible.

The $b$-flavor of the accompanying $B$ meson is identified
from inclusive properties of particles
that are not associated with the reconstructed
$\bz\to\ks\ks\ks$ candidates.
The algorithm for flavor tagging is described in detail
elsewhere~\cite{bib:fbtg_nim}.
We use two parameters, $q$ defined in Eq.~(\ref{eq:psig})
and $r$, to represent the tagging information.
The parameter $r$ is an event-by-event,
MC-determined flavor-tagging dilution factor 
that ranges from $r=0$ for no flavor
discrimination to $r=1$ for unambiguous flavor assignment.
It is used only to sort data into six $r$ intervals.
The wrong tag fractions $w$ for each of the $r$ intervals and
their differences $\Delta w$ for $\bz$ and $\bzb$ decays are
determined from data~\cite{Abe:2004xp}.

The vertex position
for $\bz\to\ks\ks\ks$ decays is
obtained using $\kspm$ trajectories and a constraint on the IP;
the IP profile
($\sigma_x\simeq100~\mu$m, $\sigma_y\simeq5~\mu$m, $\sigma_z\simeq3$~mm)
is convolved with finite $\bz$ flight length in the plane
perpendicular to the $z$ axis.
To reconstruct the $\kspm$ trajectory
with sufficient resolution,
both charged pions from the $\ks$ decay are required to
have enough SVD hits;
at least one layer with hits on both sides and at least
one additional $z$ hit in other layers for SVD-I, and
at least two layers with hits on both sides for SVD-II.
The reconstruction efficiency depends both on
the $\kspm$ momentum and on the SVD geometry.
The vertex efficiencies with SVD-II
(86\% for $\kspm\kspm\kspm$ and 74\% for $\kspm\kspm\kszz$)
are higher than those with SVD-I
(79\% for $\kspm\kspm\kspm$ and 62\% for $\kspm\kspm\kszz$)
because of the larger outer radius and the additional detector layer.
The typical vertex resolution is
about 97~$\mu$m (113~$\mu$m) for SVD-I (SVD-II) when two or three
$\kspm$ candidates can be used.
The resolution is worse when only one $\kspm$ can be used;
the typical value is 152~$\mu$m (168~$\mu$m) for SVD-I (SVD-II),
which is comparable to the $\ftag$ vertex resolution.
The determination of the vertex of the $\ftag$ final state
is the same as the $\bz\to\phi\ks$ analysis,
and is described in detail elsewhere~\cite{Abe:2004xp,bib:resol_nim}.


We determine $\cals$ and $\cala$ by performing an unbinned
maximum-likelihood fit to the observed $\Dt$ distribution.
The probability density function (PDF) expected for the signal
distribution, ${\cal P}_{\rm sig}(\Dt;\cals,\cala,\fq,w,\Delta w)$, 
is given by Eq.~(\ref{eq:psig}) after incorporating
the effect of incorrect flavor assignment.
The distribution is convolved with the
proper-time interval resolution function, $\rsig$,
which is a function of event-by-event vertex errors.


We find from MC simulation that universal $\rsig$ parameters
used for measurements of $CP$ asymmetries
in the $\bz\to\jpsi\ks$ and related 
decays~\cite{bib:BELLE-CONF-0436,bib:resol_nim}
approximately describe the resolution for the
$\bz\to\ks\ks\ks$ decay.
To account for differences between $\kspm$ trajectories and charged
tracks, additional parameters that rescale vertex errors
are introduced.
When only one $\kspm$ is used in the vertex fit,
these parameters 
are determined from a fit to
the $\Dt$ distribution of $\bz\to\jpsi\ks$ candidates,
where only the $\ks$ and the IP constraint are used for the
vertex reconstruction.
The procedure is the same as that for 
the $\bz\to\ks\piz$ decay
and is described elsewhere~\cite{Abe:2004xp}.
For events with two or three $\kspm$ used in the vertexing,
we also find from MC simulation that 
the resolution is well described by
the same $\rsig$ parameterization with an additional correction function
that depends on the number of $\kspm$ decays used for the vertex
reconstruction.
The form of this correction function is determined from a study using MC
simulation.

We determine the following likelihood value
for each event $i$:
\begin{eqnarray}
P_i
&=& (1-\fol)\int \biggl[
\fsig{\cal P}_{\rm sig}(\Dt')R_{\rm sig}(\Dt_i-\Dt') \nonumber \\
&+&(1-\fsig){\cal P}_{\rm bkg}(\Dt')R_{\rm bkg}(\Dt_i-\Dt')\biggr]
d(\Dt') \nonumber \\
&+& \fol P_{\rm ol}(\Dt_i),
\label{eq:likelihood}
\end{eqnarray}
where $P_{\rm ol}$ is a broad Gaussian function that represents
an outlier component with a small fraction $\fol$~\cite{bib:resol_nim}.
The signal probability $\fsig$
is calculated on an event-by-event basis from the function
obtained by the $\dE$-$\mb$ two-dimensional fit used to extract the
signal yield.
A PDF for background events, ${\cal P}_{\rm bkg}$,
is modeled as a sum of exponential and prompt components, and
is convolved with a sum of two Gaussians $R_{\rm bkg}$.
All parameters in ${\cal P}_{\rm bkg}$
and $R_{\rm bkg}$ are determined by the fit to the $\Dt$ distribution of a 
background-enhanced control sample, i.e. events 
outside of the $\dE$-$\mb$ signal region.
We fix $\tau_\bz$ and $\dmd$ at their world-average
values~\cite{bib:PDG2004}.
In order to reduce the statistical error on $\cala$,
we include events without vertex information.
The likelihood value in this case is obtained from the function of
Eq.~(\ref{eq:likelihood}) integrated over $\Dt_i$.

\begin{figure}
\resizebox{!}{0.84\columnwidth}{\includegraphics{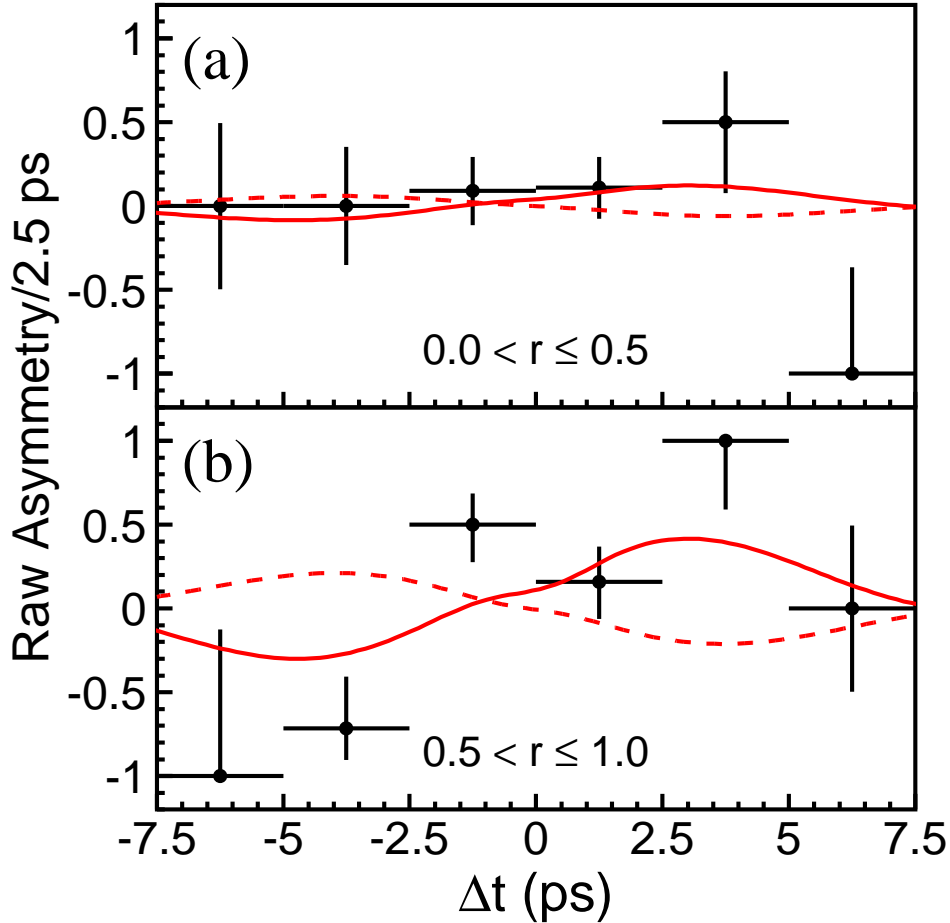}} 
\caption{
Raw asymmetry 
in each $\Dt$ bin with (a) $0 < r \le 0.5$ and (b) $0.5 < r \le 1.0$.
The solid curves show the result of the 
unbinned maximum-likelihood fit.
The dashed curves show the SM expectation with 
$(\cals, \cala) = (-0.73, 0)$.
The numbers of candidate events used for the $\cals$ measurement
are 65 for (a) $0 < r \le 0.5$ and 52 for (b) $0.5 < r \le 1.0$.
}
\label{fig:asym}
\end{figure}

The only free parameters in the final fit
are $\cals$ and $\cala$, which are determined by maximizing the
likelihood function
$L = \prod_iP_i(\Dt_i;\cals,\cala)$
where the product is over all events.
An unbinned maximum likelihood fit to the
167 $\bz\to\ks\ks\ks$ candidates, 
containing $\NBsigksksks$ $\ks\ks\ks$ signal events, yields
  \begin{eqnarray}
    \cals     &=& \SksksksResultSS,\nonumber \\
    \cala     &=& \AksksksResultSS.\nonumber
  \end{eqnarray}
We define the raw asymmetry in each $\Dt$ interval by
$(N_{+}-N_{-})/(N_{+}+N_{-})$,
where $N_{+(-)}$ is the number of 
observed candidates with $q=+1(-1)$.
%
The raw asymmetries in two regions of the flavor-tagging parameter $r$
are shown in Fig.~\ref{fig:asym}.
Note that these are simple projections onto the $\Delta t$ axis 
and do not reflect other event-by-event information (such as the signal
fraction, the wrong tag fraction and the vertex resolution), which is in
fact used in the unbinned maximum-likelihood fit for $\cals$ and $\cala$.

The systematic error is primarily due to
the resolution function
($\pm 0.12$ for $\cals$ and $\pm 0.04$ for $\cala$),
the background fractions ($\pm 0.10$ for $\cals$ and $\pm 0.03$ for $\cala$),
fit bias ($\pm 0.08$ for $\cals$ and $\pm 0.05$ for $\cala$), and 
background modeling ($\pm 0.08$ for $\cals$ and $\pm 0.01$ for $\cala$).
Other sources of systematic error are uncertainties in the wrong tag
fraction ($\pm 0.04$ for $\cals$ and $\pm 0.01$ for $\cala$),
physics parameters $\Delta m_d$ and $\tau_{B^0}$
($\pm 0.01$ for $\cals$ and $\pm 0.01$ for $\cala$),
the vertexing ($\pm 0.02$ for $\cals$ and $\pm 0.05$ for $\cala$),
and the effect of tag side interference~\cite{Long:2003wq}
($\pm 0.02$ for $\cals$ and $\pm 0.02$ for $\cala$).
We add each contribution in quadrature to obtain the total systematic errors.

Various cross-checks of the measurement are performed.
We reconstruct $\bpm\to\ks\ks\kpm$ decays
without using the charged kaon for the vertex reconstruction
and apply the same fit procedure.
We obtain
$\cals_{\ks\ks\kpm} = \SkskskpmResultS$ and
$\cala_{\ks\ks\kpm} = \AkskskpmResultS$,
which are consistent with no $CP$ asymmetry.
MC pseudo-experiments are generated to
perform ensemble tests.
We find that the statistical errors obtained
in our measurement are all consistent
with the expectations from the ensemble tests.
We apply the same procedure to the $\bz\to\jpsi\ks$ sample
without $\jpsi$ daughter tracks for vertex reconstruction.
We obtain
$\cals_{\jpsi\ks} = +0.68\pm 0.10$(stat) and
$\cala_{\jpsi\ks} = +0.02\pm 0.04$(stat), which
are in good agreement with the world average values~\cite{bib:HFAG}.
We conclude that the vertex resolution for the $\bz\to\ks\ks\ks$ decay
is well understood.

We use a frequentist approach~\cite{FeldmanCousins}
to determine the statistical significance of the deviation from the SM.
From 1-dimensional confidence intervals for $\cals$, the case with
$\cals = -0.73$ for $\bz\to\ks\ks\ks$ is ruled out at a 99.7\%
confidence level, equivalent to $2.9\sigma$ significance for Gaussian
errors.

In summary, we have performed the measurement of 
$CP$-violation parameters
in the  $\bz \to \ks\ks\ks$ decay
based on a sample of $275\times 10^6$ $B\bbar$ pairs.
The decay is dominated by the $b\to s$ flavor-changing
neutral current and the $\ks\ks\ks$ final state is a $CP$ eigenstate.
It is therefore sensitive to a possible new $CP$-violating phase
beyond the SM.
The result differs from the SM expectation by 2.9 standard deviations.


We thank the KEKB group for the excellent operation of the
accelerator, the KEK cryogenics group for the efficient
operation of the solenoid, and the KEK computer group and
the NII for valuable computing and Super-SINET network
support.  We acknowledge support from MEXT and JSPS (Japan);
ARC and DEST (Australia); NSFC (contract No.~10175071,
China); DST (India); the BK21 program of MOEHRD and the CHEP
SRC program of KOSEF (Korea); KBN (contract No.~2P03B 01324,
Poland); MIST (Russia); MHEST (Slovenia);  SNSF (Switzerland); NSC and MOE
(Taiwan); and DOE (USA).

{\it Note added.}---As we were preparing to submit this
paper, we became aware of a paper
from the BaBar collaboration~\cite{Aubert:2005dr}
which reports on the branching fraction and
$CP$ asymmetries in the $\bz\to\ks\ks\ks$ decay.



\begin{thebibliography}{999}
\bibitem{bib:CC}
Throughout this paper,
the inclusion of the charge conjugate decay mode is implied
unless otherwise stated.
\end{thebibliography}
\end{document}